%% file: main.tex
\documentclass{article}
\usepackage{spconf,amsmath,amssymb,graphicx,url}
\usepackage{tikz}
\usepackage{comment}
\usepackage{ascmac}
\usepackage{subfigure}
\usepackage{colortbl}
\usepackage[
    backend=biber,
    style=ieee,
    maxbibnames=3,
    maxcitenames=3,
    doi=false,isbn=false,url=false,eprint=false
]{biblatex} 

\input{local_definition}

\title{J-MAC: Japanese multi-speaker audiobook corpus for speech synthesis}
\name{Shinnosuke Takamichi, Wataru Nakata, Naoko Tanji, Hiroshi Saruwatari\thanks{This work is also supported by JSPS KAKENHI 19H01116, 21H04900, and JST, Moonshot R\&D Grant Number JPMJPS2011.}}
\address{The University of Tokyo, Japan}

\begin{document}
\ninept
\maketitle
\setlength{\tabcolsep}{1.1mm} 

\begin{abstract}
    In this paper, we construct a Japanese audiobook speech corpus called ``J-MAC'' for speech synthesis research. With the success of reading-style speech synthesis, the research target is shifting to tasks that use complicated contexts. Audiobook speech synthesis is a good example that requires cross-sentence, expressiveness, etc. Unlike reading-style speech, speaker-specific expressiveness in audiobook speech also becomes the context. To enhance this research, we propose a method of constructing a corpus from audiobooks read by professional speakers. From many audiobooks and their texts, our method can automatically extract and refine the data without any language dependency. Specifically, we use vocal-instrumental separation to extract clean data, connectionist temporal classification to roughly align text and audio, and voice activity detection to refine the alignment. J-MAC is open-sourced in our project page. We also conduct audiobook speech synthesis evaluations, and the results give insights into audiobook speech synthesis.
\end{abstract} 

\input{sec/introduction}
\input{sec/corpus}
\input{sec/experiment}
\input{sec/conclusion}

\printbibliography

\end{document}

%% file: local_definition.tex
\newcommand{\Fig}[1]{Figure~\ref{fig:#1}} 
\newcommand{\Table}[1]{Table~\ref{tab:#1}} 

\newcommand{\drawfig}[4]{ 
    \begin{figure}[#1]
    \centering 
    \vspace{0mm}
    \includegraphics[width=#2,clip]{#3.pdf} 
    \vspace{-3mm}
    \caption{#4}
    \label{fig:#3}
    \vspace{-2mm}
    \end{figure}
}

\newcommand{\graybox}[1]{\cellcolor[gray]{0.8}#1}

\DeclareSourcemap{
	\maps[datatype=bibtex, overwrite=true]{
		\map{
			\step[fieldsource=booktitle,
			match=\regexp{.*Interspeech.*},
			replace={Proc. Interspeech}]
			\step[fieldsource=journal,
			match=\regexp{.*INTERSPEECH.*},
			replace={Proc. Interspeech}]
			\step[fieldsource=booktitle,
			match=\regexp{.*ICASSP.*},
			replace={ICASSP}]
			\step[fieldsource=booktitle,
			match=\regexp{.*icassp_inpress.*},
			replace={ICASSP (in press)}]
			\step[fieldsource=booktitle,
			match=\regexp{.*International.*Conference.*on.*Acoustics,.*Speech.*and.*Signal.*Processing.*},
			replace={ICASSP}]
			\step[fieldsource=booktitle,
			match=\regexp{.*International.*Conference.*on.*Learning.*Representations.*},
			replace={ICLR}]
			\step[fieldsource=booktitle,
			match=\regexp{.*International.*Conference.*on.*Machine.*Learning.*},
			replace={ICML}]
			\step[fieldsource=booktitle,
			match=\regexp{.*Automatic.*Speech.*Recognition.*and.*Understanding.*},
			replace={Proc. ASRU}]
			\step[fieldsource=booktitle,
			match=\regexp{.*Spoken.*Language.*Technology.*},
			replace={Proc. SLT}]
			\step[fieldsource=booktitle,
			match=\regexp{.*Speech.*Synthesis.*Workshop.*},
			replace={Proc. SSW}]
			\step[fieldsource=booktitle,
			match=\regexp{.*workshop.*on.*speech.*synthesis.*},
			replace={Proc. SSW}]
			\step[fieldsource=booktitle,
			match=\regexp{.*Advances.*in.*neural.*information.*processing.*},
			replace={Proc. NIPS}]
			\step[fieldsource=booktitle,
			match=\regexp{.*Advances.*in.*Neural.*Information.*Processing.*},
			replace={Proc. NIPS}]
			\step[fieldsource=booktitle,
			match=\regexp{.*Workshop.*on.* Applications.* of.* Signal.*Processing.*to.*Audio.*and.*Acoustics.*},
			replace={Proc. WASPAA}]
			\step[fieldsource=booktitle,
			match=\regexp{.*International.*Conference.*on.*Language.*Resources.*and.*Evaluation.*},
			replace={Proc. LREC}]
			\step[fieldsource=publisher,
			match=\regexp{.+},
			replace={{}}]
			\step[fieldsource=month,
			match=\regexp{.+},
			replace={{}}]
			\step[fieldsource=location,
			match=\regexp{.+},
			replace={{}}]
			\step[fieldsource=address,
			match=\regexp{.+},
			replace={{}}]
			\step[fieldsource=organization,
			match=\regexp{.+},
			replace={{}}]
			\step[fieldsource=doi,
			match=\regexp{.+},
			replace={{}}]
			\step[fieldsource=url,
			match=\regexp{.+},
			replace={{}}]
			\step[fieldsource=editor,
			match=\regexp{.+},
			replace={{}}]
		}
	}
} 

%% file: sec/introduction.tex
\vspace{-3mm}
\section{Introduction} \vspace{-2mm}\label{sec:introduction}
Due to the development of deep learning, significant progress has been made on text-to-speech synthesis.
Self and source-target attention~\cite{shen18tacotron2,li19transformertts}, duration intervention~\cite{ren2021fastspeech,kim21vits}, adversarial training~\cite{saito18advss,binkowski19gantts}, and score matching~\cite{chen21wavegrad2,kong21diffwave} have enabled us to synthesize high-fidelity (near-human quality) reading-style speech. In response to these developments, the input for speech synthesis is shifting from simple textual context to more complicated contexts, e.g., dialogue~\cite{guo21conversationalend2endtts} and emotion~\cite{raitio20controllable,zhu19controlling}.

Audiobook speech synthesis requires the use of complicated contexts~\cite{simon18blizzardchallenge,xu21crossutterancebert,pan21chapterunderstanding}. To synthesize natural speech as an audiobook, we need to consider several contexts, which is not considered in a simple text reading task. These are, for example, cross-sentence contexts~\cite{xu21crossutterancebert,nakata21_ssw,pan21chapterunderstanding}, style~\cite{szekely12evaluatingaudiobook}, and expressiveness~\cite{simon18blizzardchallenge}. The speaker who reads the book is also an important context. For example, amateur speaker's voices tend to be less expressive than professional speakers' voices~\cite{szekely12amateuraudiobook,jung20pitchtron}. Even if different professional speakers read the same book, their language understanding and voice expression differ. Therefore, a corpus consisting of multiple professional speakers' voice should be constructed and open-sourced to accurately evaluate the performance of audiobook speech synthesis. However, it is very time-consuming to create such a corpus from scratch.

In this paper, we propose a method of constructing a multi-speaker audiobook corpus from audiobook products, and then construct a Japanese speech corpus ``J-MAC'' (Japanese Multi-speaker Audiobook Corpus) by using the proposed method. The method involves using source separation and connectionist temporal classification (CTC)~\cite{ctcsegmentation} for estimating data cleanliness. Iterative text-audio alignment and voice activity detection (VAD)-based refinement improve the text-audio fitting. We conduct audiobook speech synthesis evaluation using J-MAC and discuss what problems should be solved in the future. The contributions of this study are as follows:
    \vspace{-1mm}
    \begin{itemize} \leftskip -5mm \itemsep -0mm
        \item We construct a new audiobook corpus that includes multiple professional speakers for Japanese audiobook speech synthesis. Our corpus is open-sourced in our project page\footnote{\url{https://sites.google.com/site/shinnosuketakamichi/research-topics/j-mac_corpus}}.
        \item The evaluation results give insights into audiobook speech synthesis, e.g., 1) improving a synthesis method enhances naturalness of synthetic speech regardless of the speaker, and 2) the effects of the synthesis method, speaker, and book on naturalness are strongly entangled.
    \end{itemize}
    \vspace{-2mm}

%% file: sec/corpus.tex
\vspace{-2mm}
\section{Corpus construction} \vspace{-2mm} \label{sec:data_collection}
    We describe the steps to build the corpus, i.e., data collection, data cleansing, and alignment.

    \vspace{-2mm}
    \subsection{Data collection} \vspace{-2mm}
        We first collect audiobooks read by professional speakers. We set the following two conditions; the audiobook must have
        \begin{itemize} \leftskip -5mm \itemsep 0mm
            \item \textbf{Reference text:} We choose out-of-copyright books (novels) available on the Web and search the audiobook versions of these books. Another way of doing this is choosing the audiobooks and transcribing text by automatic speech recognition (ASR). However, we did not choose this another way because novels have many named entities that are difficult to transcribe with current ASR.
            \item \textbf{Another version by a different speaker:} Even if the same book is read by different speakers, each speaker expresses it differently. Therefore, we prioritize the collection of audiobooks by different speakers over different audiobooks by a single speaker.
        \end{itemize}
        
        Since neighboring sentences and hierarchical information help in conditioning audiobook speech synthesis models~\cite{xu21crossutterancebert,nakata21_ssw,pan21chapterunderstanding}, we create structured texts from the reference text. An example is below. The text has levels, i.e., chapter, paragraph, style, and sentence. ``Style'' means narrative or spoken (i.e., character-acting) sentences. The timings of each sentence are retrieved using the method described in the following subsection.
    
        \input{tab/kumo}
    
    \vspace{-2mm}
    \subsection{SNR-based cleansing} \vspace{-2mm} 
        An audiobook does not always have clean data. We classify the data into ``clean'', ``music-inserted'', and ``music-overlapped,'' as shown in \Fig{fig/snr2}\footnote{We did not evaluate the recording quality because We used commercial audiobook products that seem to have good recording quality.}. The first two can be used for constructing a corpus, but the last cannot. To eliminate ``music-overlapped,'' we first use a pre-trained model of single-channel vocal-instrumental separation (e.g., \cite{luo19convtasnet,hennequin19spleeter}) to separate voice and other sounds. We then apply signal processing-based VAD to the voice and calculate the signal-to-noise ratio (SNR) in the voice region, where signal and noise are voice and the other sounds, respectively. Audiobooks with worse SNR are eliminated\footnote{In the ``music-overlapped'' case, the extracted voice can be used as a corpus. However, to make our open-source corpus available to users without any pre-processing, we exclude the ``music-overlapped'' case in this paper.}.
    
        \drawfig{t}{0.75\linewidth}{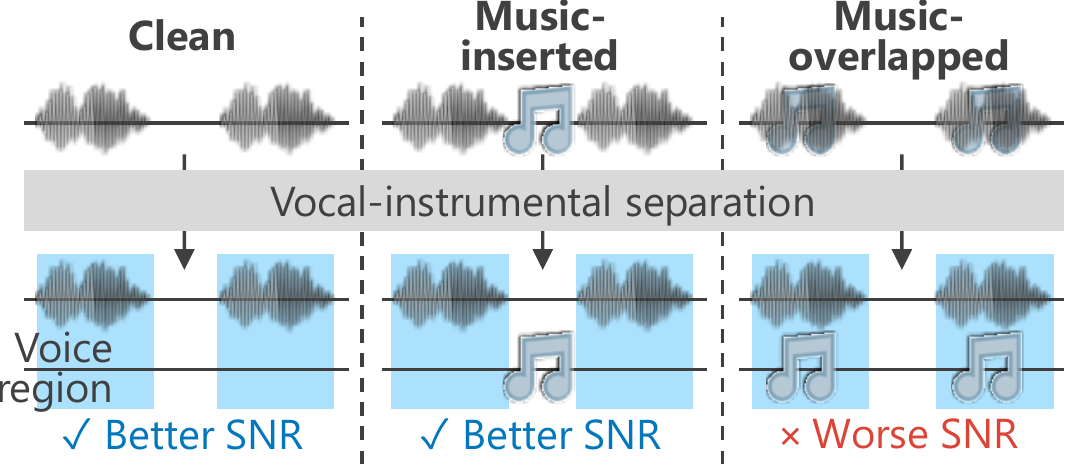}{SNR-based cleansing. SNRs are calculated in voice region estimated by VAD, and audiobooks with worse SNR are filtered out. Audiobooks that include chime sound and opening/closing music will be ``music-inserted,'' and those that include background music will be ``music-overlapped.''}
    
    \vspace{-2mm}
    \subsection{CTC-based alignment and cleansing} \vspace{-2mm}
        After SNR-based cleansing, we have pairs of structured texts and audio. We retrieve sentence-level alignments from these pairs. We use CTC segmentation~\cite{ctcsegmentation} using pre-trained end-to-end alignment models\footnote{``End-to-end model'' in this paper means that it uses tokens that do not require language-specific pre-processing, e.g., words or subwords.}. CTC segmentation aligns the sentences and audio and calculates their fitting scores\footnote{\url{https://github.com/lumaku/ctc-segmentation}}. It uses CTC log-posterior probabilities to determine sentence timings in the audio given a ground-truth text. Using an end-to-end model enables alignment without language-dependent preprocessing. However, the accuracy of the alignment is not sufficient, so the following two processes are carried out.
        
        \drawfig{t}{0.95\linewidth}{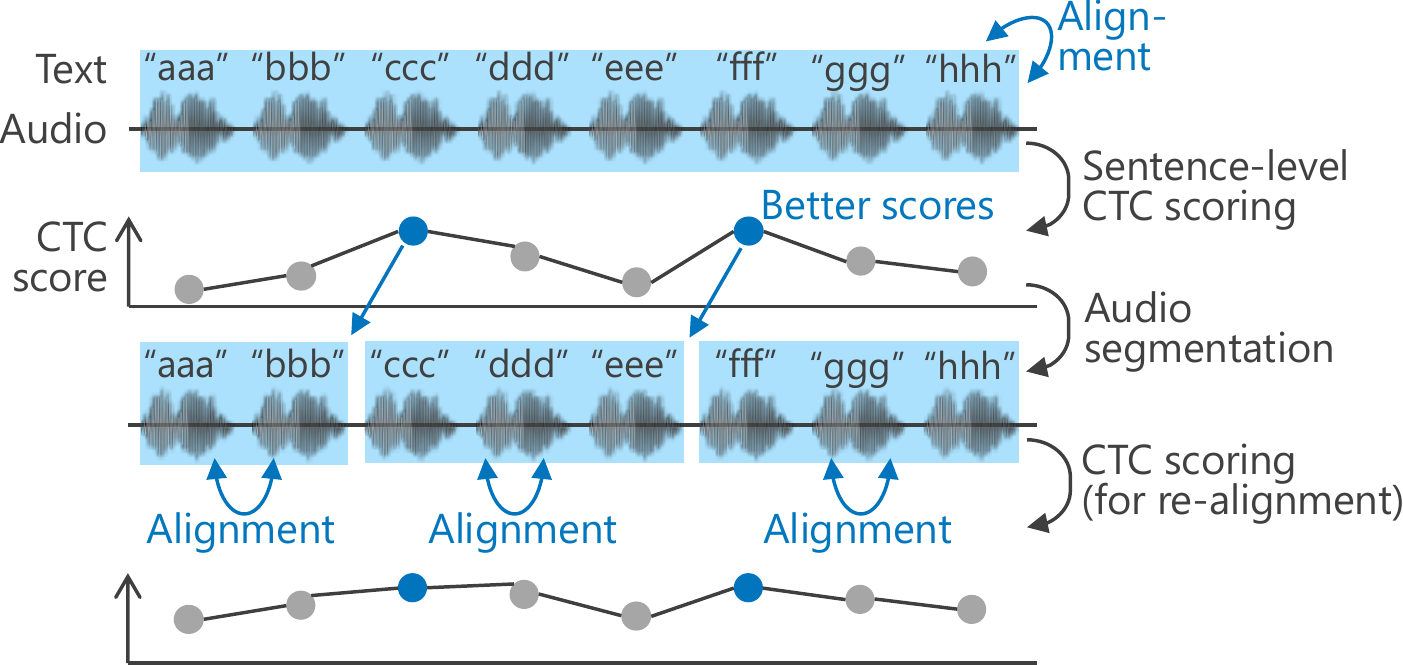}{CTC-based recursive text-audio alignment. By splitting audio using sentences with best CTC scores as delimiters, we repeat alignment to improve CTC score.}
        
        \drawfig{t}{0.65\linewidth}{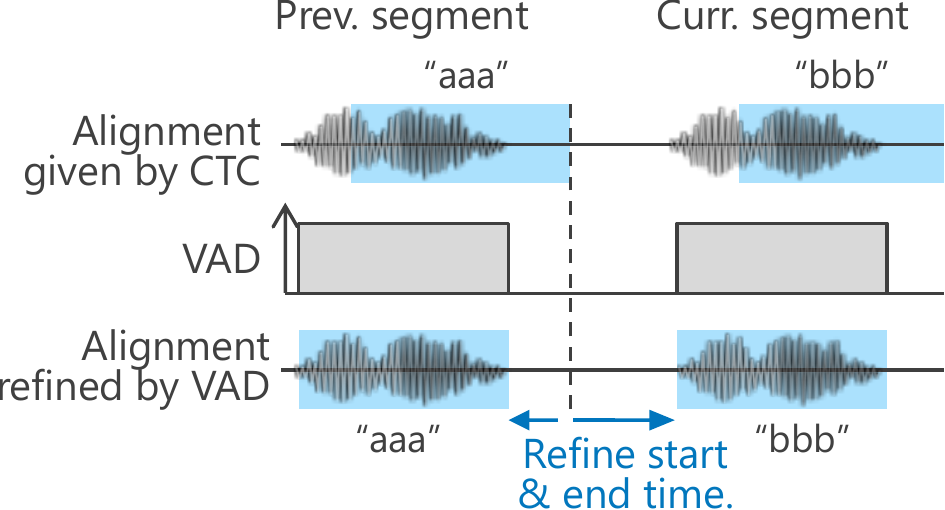}{VAD-based refinement of text-audio alignment. This eliminates offset often seen in CTC-based alignment. In practice, final alignment is range of voice segment plus small margin.}    
        
        \textbf{CTC-based re-alignment.} The duration of audiobooks ranges up to several hours. When the alignment model is trained using short audio, the alignment of such long audio tends to fail. Therefore, we re-align the long audio by dynamically splitting the audio, as shown in \Fig{fig/align}. Specifically, we first calculate the CTC score of each sentence in the first alignment, and find sentences with the best CTC scores. The voice is then split up at the aligned timings of those sentences, and alignment is carried out again on each of the split voices. This is repeated until the average CTC score of the audiobook no longer improves.

        \textbf{VAD-based refinement.} The CTC criterion does not capture the exact alignment. Therefore, CTC-based alignment tends to result in an offset in the aligned timing from the actual voice segments\footnote{This issue is discussed in \url{https://github.com/espnet/espnet/issues/3018}.}. To solve this, we use VAD to refine the timings, as shown in \Fig{fig/vad}. First, we run VAD around the end timing of each sentence to find the neighboring voice segments. We then move the end timing until it corresponds to the speech segment. Similarly, we modify the start timing of the next sentence, using the obtained voice segment. In practice, a small timing margin is added to the final alignment.

%% file: tab/kumo.tex
\begin{itembox}[l]{\footnotesize kumo.yaml (The beginning of ``The Spider's Thread'')}
\vspace{-3mm}
\footnotesize
\begin{verbatim}
chapt000:       # chapter index(000-) 
 parag000:      # paragraph index (000-)
  style000:     # style index (000-)
   time:
    - 0.96      # start time [sec] 
    - 3.32      # end time [sec] 
   - sent: "It happened one day." # 1st sentence
   time: ...
\end{verbatim}
\vspace{-6mm}
\end{itembox}

%% file: sec/experiment.tex
\vspace{-2mm}
\section{Experimental evaluation} \vspace{-2mm}\label{sec:experiment}
    \subsection{Evaluation in data collection} \vspace{-2mm}
        We crawled Audiobook.jp\footnote{\url{https://audiobook.jp}}, a largest audiobook provider in Japan, and prepared approximately 150 audiobooks. The reference text was crawled from Aozora Bunko\footnote{\url{https://www.aozora.gr.jp/}}, a Japanese digital library that stores out-of-copyright books. Texts of the books are provided in HTML format, and some characters (e.g., Chinese characters used in Japanese (kanji)) include reading aids. The structured text was automatically created from the HTML-format text. Chapter boundaries were signified with a chapter index or blank line. Boundaries of paragraphs, styles, and sentences were signified with indentations, quotation marks, and periods, respectively. The reading aids were also reflected in the structured text, e.g., kanji were accompanied with their readings. 

        \input{tab/corpus_spec}

        \Table{corpus_spec} lists the specifications of our corpus. These values are not all the collected data described above but the actual corpus developed via the following experiments. From these values, we can describe 1) $1.9$ audiobooks were spoken per speaker, 2) $3.0$ audiobooks were collected per book, and $48.0$ minutes duration were collected per speaker. \Fig{fig/audiobook_statistics} shows histograms of audiobooks per book and speaker. We could collect audiobooks read by different speakers and those read by the same speaker. 

        \drawfig{t}{0.85\linewidth}{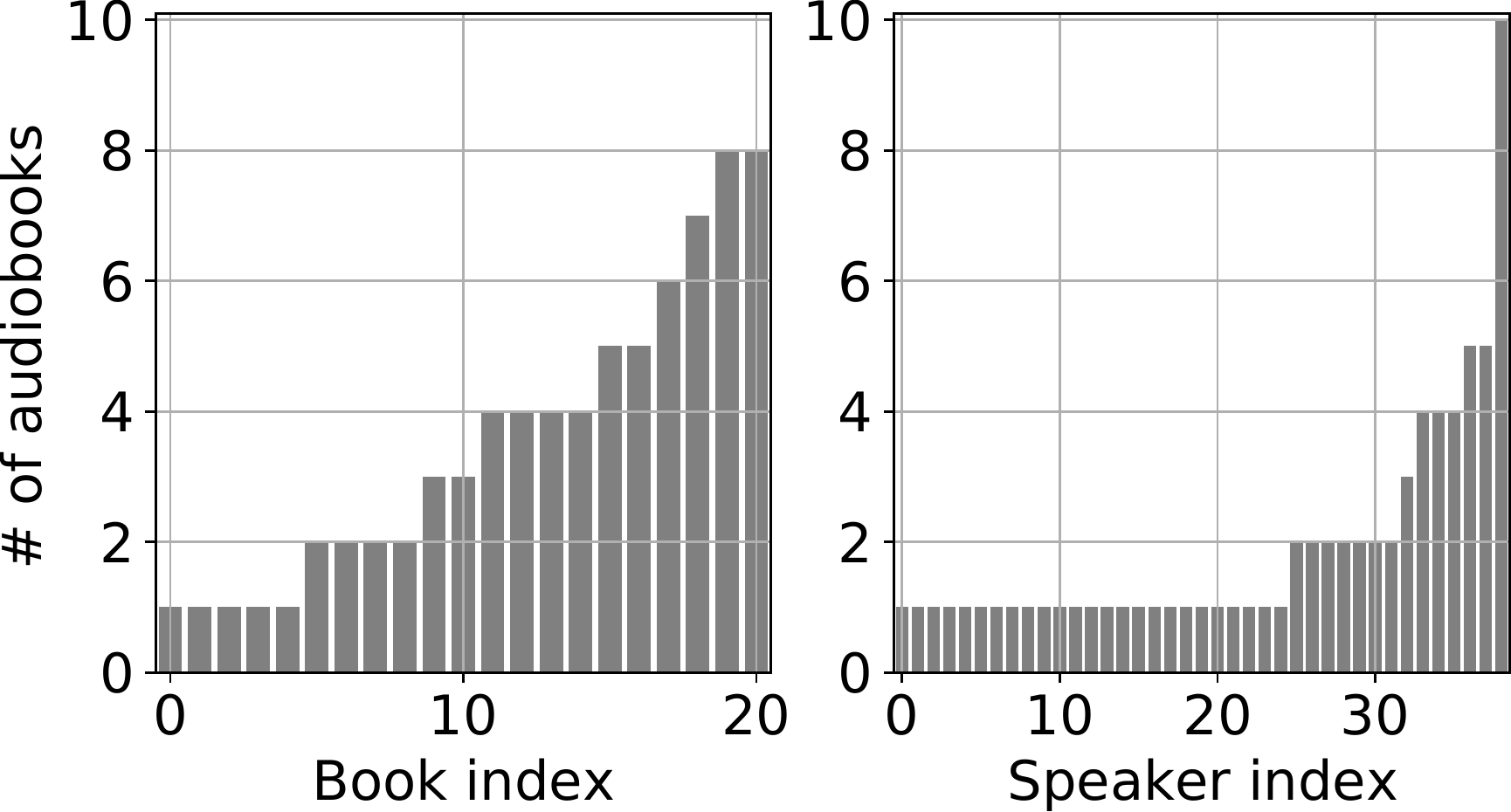}{Distribution of audiobooks per book (left) and speaker (right). Resulting corpus includes audiobooks read by different speakers and speaker who read multiple books.}

        \input{tab/corpus_comparison}

        \Table{corpus_comparison} compares J-MAC with the existing corpora. J-MAC is smaller than corpora of amateur speakers but has unique features: multiple professional speakers and parallel data (i.e., an audiobook spoken by different speakers). It is also the first open-source multi-speaker audiobook corpus in Japanese.
    
    \vspace{-2mm}    
    \subsection{Evaluation in background music estimation} \vspace{-2mm} 
        We used a pre-trained model of Spleeter\footnote{\url{https://github.com/deezer/spleeter}} for deep learning-based vocal-instrumental separation. We used py-webrtcvad\footnote{\url{https://github.com/wiseman/py-webrtcvad}} for signal processing-based VAD.  

        \drawfig{t}{0.7\linewidth}{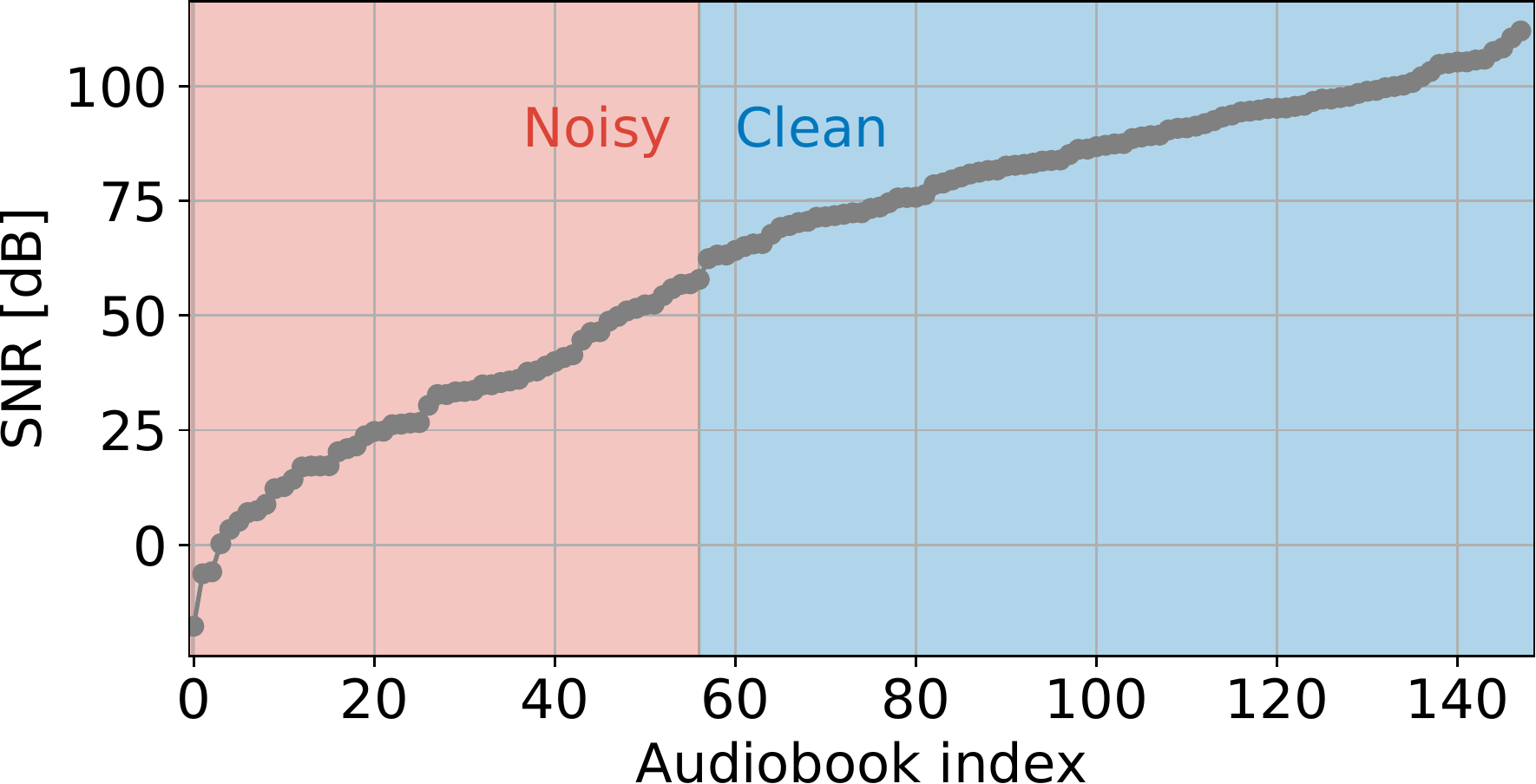}{SNR of each audiobook. Audiobooks are filtered on basis of manually determined threshold.}

        \Fig{fig/snr} shows the results of SNR calculation. The SNR ranged from $0$~dB (i.e., voices and instrumentals are similar in volume.) to $100$~dB (very clean). The threshold was determined manually because it was difficult to determine it automatically due to its gentle variation. The ``Clean'' audiobooks proceeded to segmentation, as described in the next subsection.
    
    \vspace{-2mm}
    \subsection{Evaluation in segmentation} \vspace{-2mm}
        We used an recurrent neural network (RNN)-based CTC alignment model provided by ESPnet~\cite{watanabe18espnet}. The model was trained using the CSJ corpus~\cite{maekawa00csj}. The CTC score improvement by re-alignment converged until three iterations. The $5$-best sentences on the CTC scores were used to split audio in re-alignment. 
    
        \drawfig{t}{0.7\linewidth}{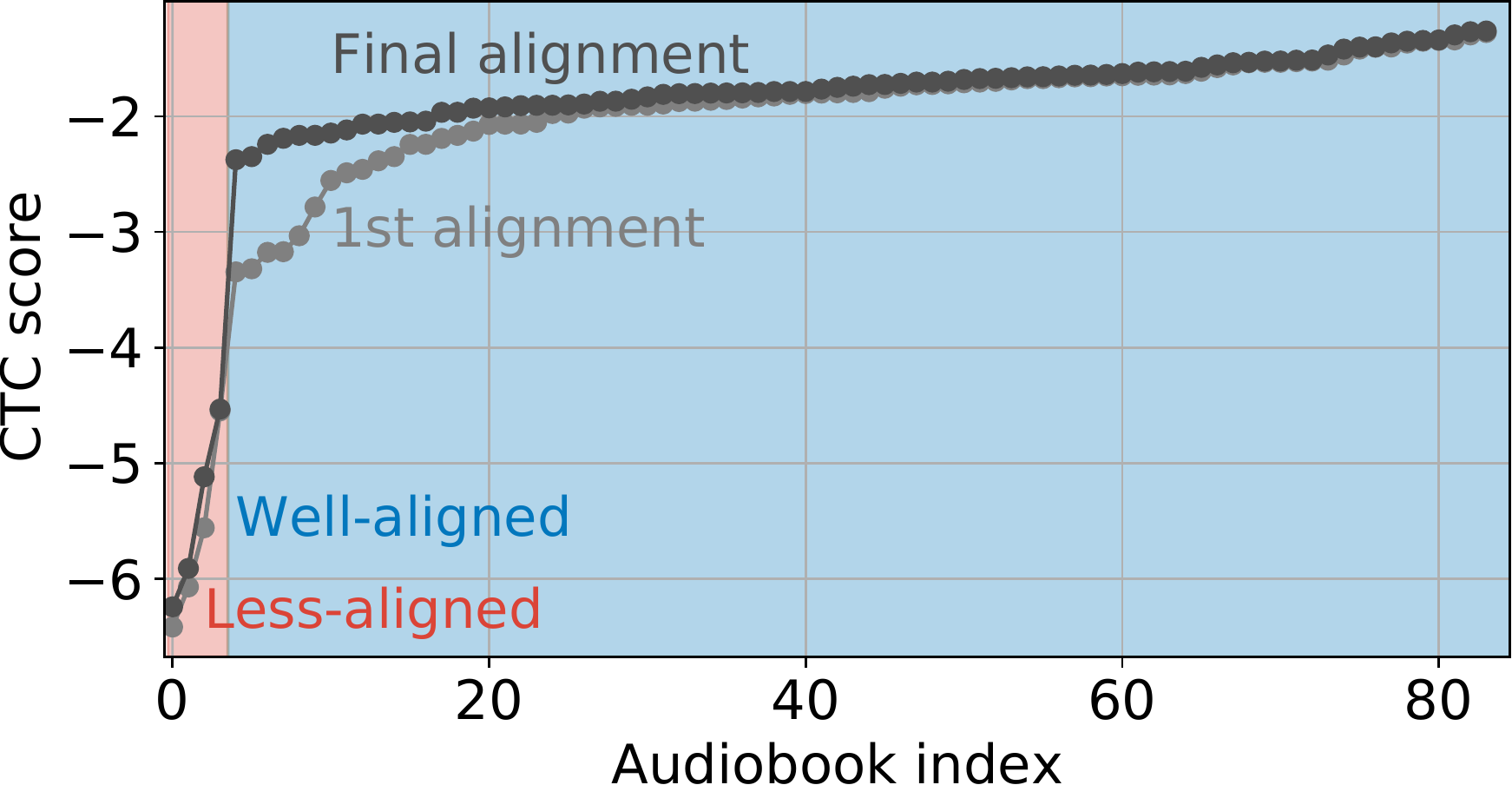}{CTC scores of each audiobook. Re-alignment improves CTC score. Audiobooks with lower CTC scores are eliminated.}
        \drawfig{t}{0.7\linewidth}{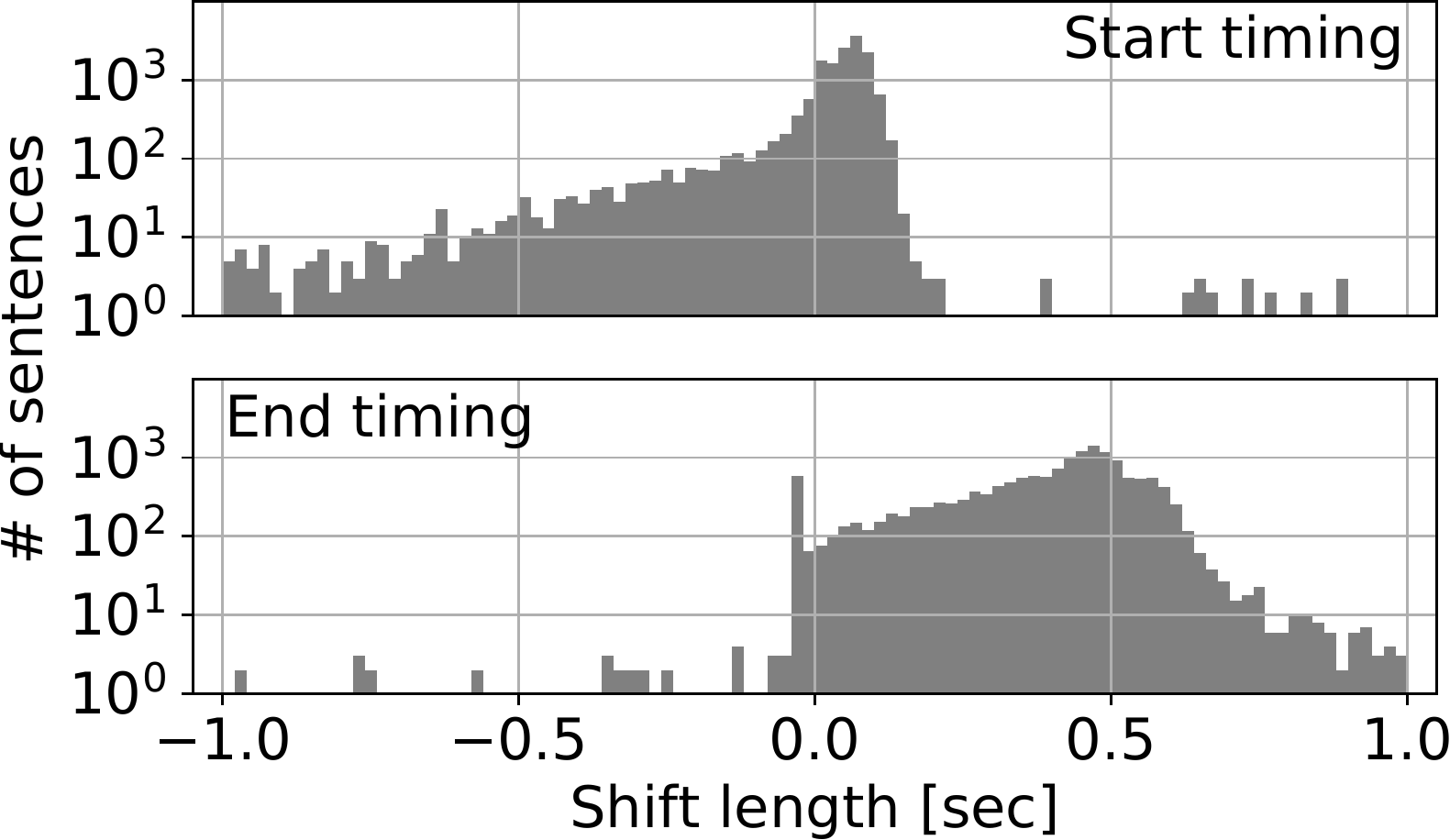}{Distribution of VAD-based timing refinement. Both start and end timings given by CTC-based alignment have offsets and are refined by VAD.}
        
        \Fig{fig/ctc} shows the CTC scores. The re-alignment significantly improved the CTC scores around $-4.0$--$-2.0$ and contributed to selecting more well-aligned audiobooks. Even after realignment, the CTC scores of certain audiobooks were significantly lower ($< -4.0$). We examined these audiobooks and found that they were translations of Japanese books into English. Although the text and audio do not basically correspond, some named entities do (e.g., readings of the main character's name in Japanese text is represented in English speech). The interesting finding is that these alignments did not fail, and CTC scores are obtained.     
        
        \Fig{fig/vad_shift} shows histograms of timing shift by VAD. Most of the aligned timings were shifted, suggesting that CTC alignment alone can result in overlapping in the speech segment; our correction using VAD avoids this problem.
        
    \vspace{-2mm}
    \subsection{Evaluation in speech synthesis} \vspace{-2mm}
        Finally, we evaluated our corpus on an audiobook speech synthesis task using multiple synthesis methods, audiobooks, and speakers. We trained a multi-speaker speech synthesis model and evaluated the synthetic speech of different books and speakers.
    
        \subsubsection{Model setup}
            
            \drawfig{t}{0.98\linewidth}{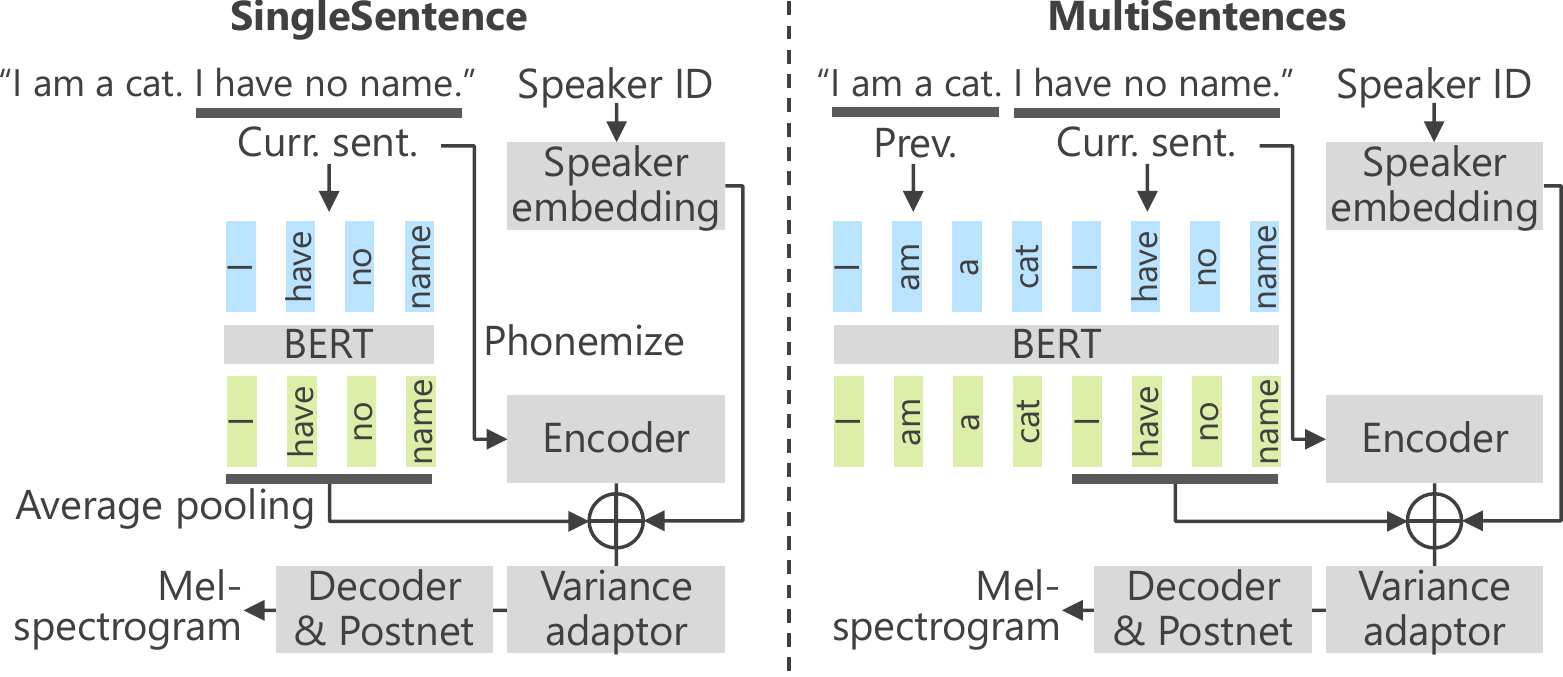}{Architectures of audiobook speech synthesis. Left considers only current sentence as well as basic text-to-speech, and right considers neighboring sentences. In practice, subsequent sentences are also taken into account, but for simplicity, only preceding sentences are input into BERT in right figure.}

            We used a modified version of Nakata's method~\cite{nakata21_ssw}, which retrieves contexts of input text using bidirectional encoder representations from Transformers (BERT)~\cite{vaswani17allYouNeed}. \Fig{fig/architecture} shows the model architecture. The ``SingleSentence'' method synthesizes speech sentence by sentence in isolation, as with the basic reading-style speech synthesis. The ``MultiSentences'' method synthesizes speech of the current sentence, using neighboring sentences (previous, current, and following sentences). The modified version of the model was based on FastSpeech2~\cite{ren2021fastspeech} instead of Tacotron2~\cite{shen18tacotron2}, and sentence-level embedding was used instead of word-level embedding. This enables MultiSentences to capture cross-sentence context while SingleSentence cannot. To train on a multi-speaker corpus, speaker lookup embedding was used. 
            
            
            We downsampled audio data to $22.05$~kHz, segmented them into a sentence level, and split them into training and validation sets in advance. The sizes of the training and validation sets were $14,043$ and $100$ samples, respectively. Julius~\cite{lee09julius} was used to obtain alignments between utterances and phoneme sequences. The generated melspectrogram configurations were 80 dimensions, with frame length of $1,024$ samples and frame shift of $256$ samples.
            
            We used the pre-trained BERT model trained using Japanese Wikipedia data. The model configuration is equivalent to BERT-tiny~\cite{turc2019wellread}. However, we did not perform pre-trained distillation. The BERT weights are unfreezed upon training except for the embedding layer as not all vocabulary appear on the training set. 
            
            For optimization, we used the Adam optimizer with $\beta_1 = 0.9$, $\beta_2 = 0.99$. Learning rate scheduling was applied in the same manner as in a previous study~\cite{vaswani17allYouNeed} with $4,000$ warmup steps, and the batch size was 64. For the loss function, we used the mean square error of each output from a variance adapter and the melspectrogram of the model output and decoder output (i.e. output before going through Postnet). The BERT weights are unfreezed upon training except for the embedding layer as not all vocabulary appear on the training set. We used HiFiGAN~\cite{kong20hifigan} with the weights distributed on official code implementation\footnote{\url{https://github.com/jik876/hifi-gan}}. Specifically, we used the UNIVERSAL\_V1 model. We did not perform finetuning of the HiFiGAN parameters.
            
        \subsubsection{Results and analysis of subjective evaluation}
            We conducted a mean opinion score (MOS) evaluation on the naturalness of audiobook speech. We explored the effect of speaker, book, and synthesis method on naturalness. Under these conditions, we synthesized audiobook speech in paragraph level containing 2--5 sentences. Six speakers (3 males and 3 females labeled ``m1''--``f3'') were randomly selected from the corpus. The test set was four audiobooks ("akai", "akazukin", "donguri", and "tsuchigami," which are labeled ``book1''--``book4,'' respectively) selected from the J-KAC corpus~\cite{nakata21_ssw}. The test set had no overlap from the training and validation sets, but all the sets belonged to the same domain of books (all were developed from the same source (Aozora Bunko)). 260 listeners participated to the MOS test and each listener evaluated randomly selected 10 samples.   
            
            \Table{mos} lists the MOS score under each condition. From these results, we can provide the following insights.
            \begin{itemize} \itemsep 0mm \leftskip -5mm 
                \item \textbf{Synthesis method}: Considering multiple sentences slightly improves naturalness ($3.24$ vs. $3.30$) in audiobook speech synthesis.
                \item \textbf{Books}: The naturalness significantly changes by book ($3.01$--$3.45$). As mentioned above, the domains of these books and the training data are the same, so factors other than the domain affect the scores. 
                \item \textbf{Speakers}: The naturalness also changes by speaker ($2.97$--$3.43$). The amount of training data for the low-scoring speakers ``m2'' and ``m3'' (2 and 4 audiobooks, respectively) is more than those for ``m1'' and ``m3'' (1 audiobook, respectively), so we cannot say that the reason for this result is the amount of training data alone.
            \end{itemize}

            \input{tab/mos}

            \Table{statistics} shows the results of the analysis of variance (ANOVA). Statistical significance was observed in each of synthesis method, book, and speaker. These results also indicate that there is an interaction among each factor, except for between method and speaker. This suggests that a synthesis method using multiple sentences improves naturalness to the same extent regardless of speaker, and that there is an entanglement between synthesis methods, speakers, and books, and that this disentanglement is necessary to improve the quality of audiobook speech synthesis. These will be good directions for audiobook speech synthesis research using this corpus.

            \input{tab/statistics}

%% file: tab/corpus_spec.tex
\begin{table}[t]
    \vspace{-3mm}
    \centering
    \caption{Corpus specification}
    \begin{tabular}{l|l}
        Term                                & Value  \\ \hline
        \# of speakers                          & 39 \\
        \# of books                             & 24 \\
        \# of audiobooks                        & 74 \\
        \# of book-authors                       & 7 \\
        Duration [hour]                     & 31.5 \\
        Sampling rate [Hz]                  & 44.1  
    \end{tabular}
    \label{tab:corpus_spec}
    \vspace{-3mm}
\end{table}


%% file: tab/corpus_comparison.tex
\begin{table}[t]
    \vspace{-3mm}
    \centering
    \footnotesize
    \caption{Comparison of open-sourced audiobook corpora. Duration in ``hours.''}
    \begin{tabular}{c|ccccc}
    Lang.               & Corpus        & Duration  & \# of speakers & Professional & Parallel \\ \hline
    En             & Blizzard2018~\cite{simon18blizzardchallenge}      & 6.5               & 1       & Yes    & No\\
    En             & Blizzard2013~\cite{simon13blizzardchallenge}      & 300               & 1         & Yes  & No \\
    En             & LibriTTS~\cite{zen19libritts}      & 585               & 2,456       & No    & Yes \\
    Fr              & SynPaFlex~\cite{sini18SynPaFlex}     & 87               & 1       & No     & No \\    
    Multi            & TUNDRA~\cite{stan13tundra}        & 60               & 14       & No & No \\
    Ja            & J-KAC~\cite{nakata21_ssw}         & 9                 & 1         & Yes & No \\
    Ja            & \textbf{J-MAC (ours)}         & 32                & 39       & \textbf{Yes}   & \textbf{Yes} \\
    \end{tabular}
    \label{tab:corpus_comparison}
    \vspace{-3mm}
\end{table}

%% file: tab/mos.tex
\begin{table}[t]
\vspace{-3mm}
\centering
\caption{MOS under several speaker (m1--f3) and book (book1--book4) conditions. ``*/*'' indicates scores of two methods: ``SingleSentence'' and ``MultiSentences.'' Gray boxes are worst and best values. ``Ave.'' rows and columns indicate average scores.}

\begin{tabular}{r|c|c|c|c|c}
     & book1 & book2 & book3 & book4 & Ave. \\ \hline
m1   & 3.20 / 3.18 & 3.53 / 3.00 & 3.23 / 3.57 & 3.13 / 3.10 & 3.21 / 3.24 \\
m2   & 3.24 / 3.35 & 3.33 / 3.38 & 3.48 / 3.27 & 2.97 / 3.08 & 3.19 / 3.23\\
m3   & 3.07 / 3.34 & 3.08 / 3.00 & 3.35 / 3.29 & \graybox{2.62 / 2.99} & 2.97 / 3.19 \\
f1   & 3.57 / 3.59 & 3.64 / \graybox{3.73 & 3.49} / 3.57 & 3.15 / 3.21 & 3.43 / 3.49\\
f2   & 3.30 / 3.45 & 3.29 / 3.00 & 3.65 / 3.29 & 3.24 / 3.35 & 3.36 / 3.35\\
f3   & 3.42 / 3.52 & 3.61 / 3.44 & 3.25 / 3.16 & 3.07 / 3.12 & 3.28 / 3.29 \\ \hline
Ave. & 3.29 / 3.41 & 3.45 / 3.29 & 3.41 / 3.36 & 3.01 / 3.14 & 3.24 / 3.30
\end{tabular}
\vspace{-3mm}
\label{tab:mos}
\end{table}

%% file: tab/statistics.tex
\begin{table}[t]
\vspace{-1mm}
\centering
\caption{ANOVA results of audiobook speech synthesis. ``$*$'' indicates interaction of two or three factors. \graybox{Gray box} indicates $p < 0.05$.}
\begin{tabular}{l|c}
Factor                       & $p$-value \\ \hline
method                       & \graybox{0.041} \\
speaker                      & \graybox{0.000} \\
book                    & \graybox{0.000} \\
method * speaker             & 0.225 \\
method * book           & \graybox{0.009} \\
speaker * book          & \graybox{0.000} \\
method * speaker * book & \graybox{0.048}
\end{tabular}
\label{tab:statistics}
\vspace{-3mm}
\end{table}

%% file: sec/conclusion.tex
\vspace{-2mm}
\section{Conclusion} \vspace{-2mm}\label{sec:conclusion}
    We proposed a method of constructing an audiobook speech corpus and constructed $31.5$ hours of data for Japanese audiobook speech synthesis. We evaluated the corpus in terms of audiobook speech synthesis and suggested factors for future direction of audiobook speech synthesis.
    
    J-MAC is open-sourced in our project page and is for research purpose only. The corpus does not include audio data, and users need to purchase the audiobook products from the websites listed in the corpus.
